\documentclass[useAMS,usenatbib]{mn2e}
\usepackage{epsfig,amssymb,amsmath,graphicx,lscape}
\usepackage[figuresright]{rotating}

\title[AstroSat view of LMC X-2]{{\it AstroSat} view of 
LMC X-2: Evolution of broadband X-ray spectral properties along a complete Z-track}
\author[V. K. Agrawal ]{V. K. Agrawal$\thanks{E-mail:
vivekag@ursc.gov.in}$, Anuj Nandi\\
Space Astronomy Group, ISITE Campus, U. R. Rao Satellite Center, Bangalore, 560037, India \\
}
\begin{document}


\pagerange{\pageref{firstpage}--\pageref{lastpage}}
\maketitle
\label{firstpage}

\begin{abstract}
In this paper, we report the first results of the extragalactic Z-source
LMC X-2 obtained using the $\sim$ 140 ks observations with  {\it Large
Area X-ray Proportional Counter (LAXPC)} and {\it Soft X-ray Telescope
(SXT)} onboard {\it AstroSat}. The HID created with the {\it LAXPC}
data revealed a complete Z-pattern of the source, showing all the three
branches. We studied the evolution of the broadband X-ray spectra in the
energy range of $0.5-20.0$ keV along the Z-track, a first such study of
this source. The X-ray spectra of the different parts of the Z-pattern
were well described by an absorbed Comptonized component. An absence of
the accretion disc component suggests that the disc is most probably
obscured by a Comptonized region. The best fit electron temperature
($kT_e$) was found to be in the range of $1.7-2.1$ keV and optical depth
($\tau$) was found to be in the range of $13.2-17.5$. The optical depth
($\tau$) increased as the source moved from the normal/flaring branch
(NB/FB) vertex to the upper part of the FB, suggesting a possible outflow
triggered by a strong radiation pressure. The power density spectra
(PDS) of HB and NB could be fitted with a pure power-law of index $\alpha$
$\sim$ 1.68 and 0.83 respectively. We also found a weak evidence of  QPO 
(2.8~$\sigma$) in the FB. The intrinsic luminosity of the source varied
between $(1.03-1.79)$ $\times$ 10$^{38}$ ergs/s. We discuss our results by
comparing with other Z-sources and the previous observations of LMC X-2.
\end{abstract}
\begin{keywords}
accretion, accretion discs - X-rays: binaries - X-rays: individual: LMC X-2
\end{keywords}

\section{Introduction}
Low mass X-ray binaries (LMXBs) are systems where a compact object accretes
matter from a low mass companion ($\leq 1~M_\odot$) via Roche lobe
overflow. LMXBs hosting a low-magnetic field neutron star provide
an ideal laboratory to study the physics of accretion processes at
the vicinity of an ultra dense compact object and in the strong gravity
regime. Early studies of the selected bright LMXBs revealed that the six
luminous LMXBs trace an approximate Z-type pattern in the Colour-Colour
Diagram (CCD) and the Hardness-Intensity Diagram (HID), and hence they were
named Z-sources \citep{Hasvan89}. The three branches of the Z-track are:
horizontal branch (HB), normal branch (NB) and flaring branch (FB). They are
the brightest X-ray binaries, accreting close to the Eddington limit
($0.5-1.0$ $L_{Edd}$). The other class of the neutron star LMXBs exhibited
a fragmented pattern in the CCD and HID, and are termed as atoll sources
\citep{Hasvan89}. Luminosities of these sources vary in a wider range
($0.01-0.2$ $L_{Edd}$).

X-ray spectra of Z-sources in the $0.5 - 30$ keV range are described by two component models. 
The soft component is either modeled by a multi-colour disc (MCD) ({\it diskbb} in XSPEC) 
emission \citep{Mit84,disalvo02,agr03, agr09} or a single temperature blackbody ({\it bbody} 
in XSPEC) \citep{disalvo00,disalvo01}. The hard component is described by a Comptonized 
component ({\it compTT} or {\it nthComp} in XSPEC), resulting from the inverse-Compton 
scattering of the soft seed photons. A combination {\it diskbb+bbody} is also frequently 
used to model the X-ray spectra of many bright LMXBs \citep{Cackett10, Lin12}. A combination 
{\it cutoff-powerlaw+bbody} is also some time used to describe the spectra of 
Z-sources \citep{balu10,Jackson09}.

Detailed studies have been carried out in order to understand
the evolution of the X-ray spectra along the Z-track (Cyg X-2:
\citealt{disalvo02, balu10, Done02, Fari09}, GX 17+2: \citealt{agr20,
disalvo00, Lin12}, GX 349+2: \citealt{agr03}; Sco X-1: \citealt{Church12};
GX 5-1: \citealt{Jackson09, Bhulla19}, GX 340+0: \citealt{Iaria06}).
Z-sources also show Quasi-periodic Oscillations (QPOs) in the range of
$5 - 1200$ Hz (for review, see \citealt{vander00}). Three types of QPOs,
horizontal branch oscillations (HBOs, $15-100$ Hz), normal/flaring branch
oscillations (N/FBOs, $5-30$ Hz) and a pair of kHz QPOs ($200-1200$ Hz)
have been reported in the Z-sources.

LMC X-2 is one of the brightest low mass X-ray binaries in the
Large Magellanic Cloud (LMC).  The optical counter-part of this
source is a variable $18^{\mathrm{th}}$  mag blue star \citep{pakul78}.
The source is reported to have persistent nature with X-ray
luminosity varying in a narrow range of $0.6-3$ $\times$ 10$^{38}$ ergs/s
\citep{Markert75,Johnston79}. Considering the high luminosity of the
source and pattern traced by it in CCD and HID, it was suggested that
LMC X-2 probably belongs to the Z-Class \citep{Smale00}. An extensive
analysis of the data from the proportional-counter-array (PCA) onboard
{\it Rossi-X-ray Timing Explorer (RXTE)} satellite revealed a complete
Z-diagram of this source, making it the first extragalactic Z-source
and seventh in this group \citep{Smale03}. The source also exhibited
8.16 hours modulation in the X-ray lightcurve \citep{Smale00}.

The {\it EXOSAT} spectra of this source in the $0.9-20$ keV band can be well fitted with either a thermal Comptonization model with temperature $\sim$ 3 keV or  a combination of a blackbody ($kT_{bb}$ $\sim$ 1.2 keV) and thermal bremsstrahlung emission ($kT_{th}$ $\sim$ 5 keV) \citep{Bonnet89}. The {\it XMM-Netwon} X-ray spectrum of the source was well described by a model consisting of a blackbody emission from the neutron star surface and disc blackbody emission from the standard thin disc \citep{lava08}.  They also attempted with a Comptonization model to describe the X-ray spectrum of the source. During their observation, the source was in the normal branch.

  \cite{agr09} carried out a
detailed spectral study of this source using {\it RXTE} and {\it Suzaku}
data. They studied the evolution of the X-ray spectra along the complete
Z-track using $3-20$ keV {\it RXTE-PCA}  observations. They fitted
the $3-20$ keV spectra with absorbed {\it compTT} model and found that
the Comptonized component comes from an optically thick (optical depth
$\tau$ $\sim$ 12) and a cool corona (electron temperature $kT_e \sim$
$1.9-2.7$ keV). They also suggested that a systematic variation in the
Compton parameter $y$ is responsible for the motion of the source along
the Z-track. They also fitted the {\it Suzaku} XIS + PIN data for both
flaring and quiet state with  two component models. They found that the
combination of a disc blackbody and {\it compTT} components provides a 
better fit compared to the  {\it bbody+compTT} model.

Till now no clear evidence of QPO feature has been found in this
source. The power density spectra (PDS) of this source for all the
three branches (HB, NB and FB) can be described by a simple power-law
\citep{Smale03}.

In this work, we present the results of the first {\it AstroSat} observations of the source LMC X-2. We have investigated the X-ray spectral evolution of the source LMC X-2 in the energy range of $0.5-20.0$ keV along the complete Z-pattern in the HID. This is the first such  detailed study of this source. We have studied the evolution of the PDS along the three branches of the Z-pattern. The remainder of the paper is organized as follows. The observations and procedure of the data reduction is presented in $\S$2. Method of data analysis and the modelling of the energy spectra and the power density spectra are presented in $\S$3. The results of the spectral and temporal analysis are described in $\S$4. Finally, we discuss the implications of our results and conclude in $\S$5.

\section{Observations and Data Reduction}

{\it AstroSat} observed the source LMC X-2 from June 22, 2016 to June 24, 2016
using the instruments ({\it SXT} and {\it LAXPC}) onboard {\it AstroSat}, for a total
exposure time of 140 ks. {\it AstroSat} provides an unique opportunity to
understand the spectral and timing behaviour of a celestial source in
$0.5-100$ keV with its suite of three co-aligned instruments: Soft X-ray
Telescope ({\it SXT}, \citet{singh16}), Large Area X-ray Proportional Counter
({\it LAXPC}, see \citet{Yadav16}) and Cadmium-Zinc-Telluride Imager ({\it CZTI}, 
\citet{vadawale16}). Here, we have used the data collected with the {\it SXT} and {\it LAXPC} 
instruments. {\it SXT} operates in 
the energy range $0.3-8.0$ keV and {\it LAXPC} operates in the $3.0-80.0$ keV band. During our 
observation, {\it SXT} was operated in the photon counting (PC) mode. The time resolution in this 
mode is 2.37~s. During our observation, {\it LAXPC} was operated in the event analysis
mode in which events were tagged with an accuracy of 10 micro-seconds.
{\it LAXPC} consists of three co-aligned identical X-ray proportional counters
(LAXPC10, LAXPC20 and LAXPC30) with combined effective area of $\sim$
6000 cm$^2$.

\begin{figure}
\includegraphics[width=0.55\textwidth]{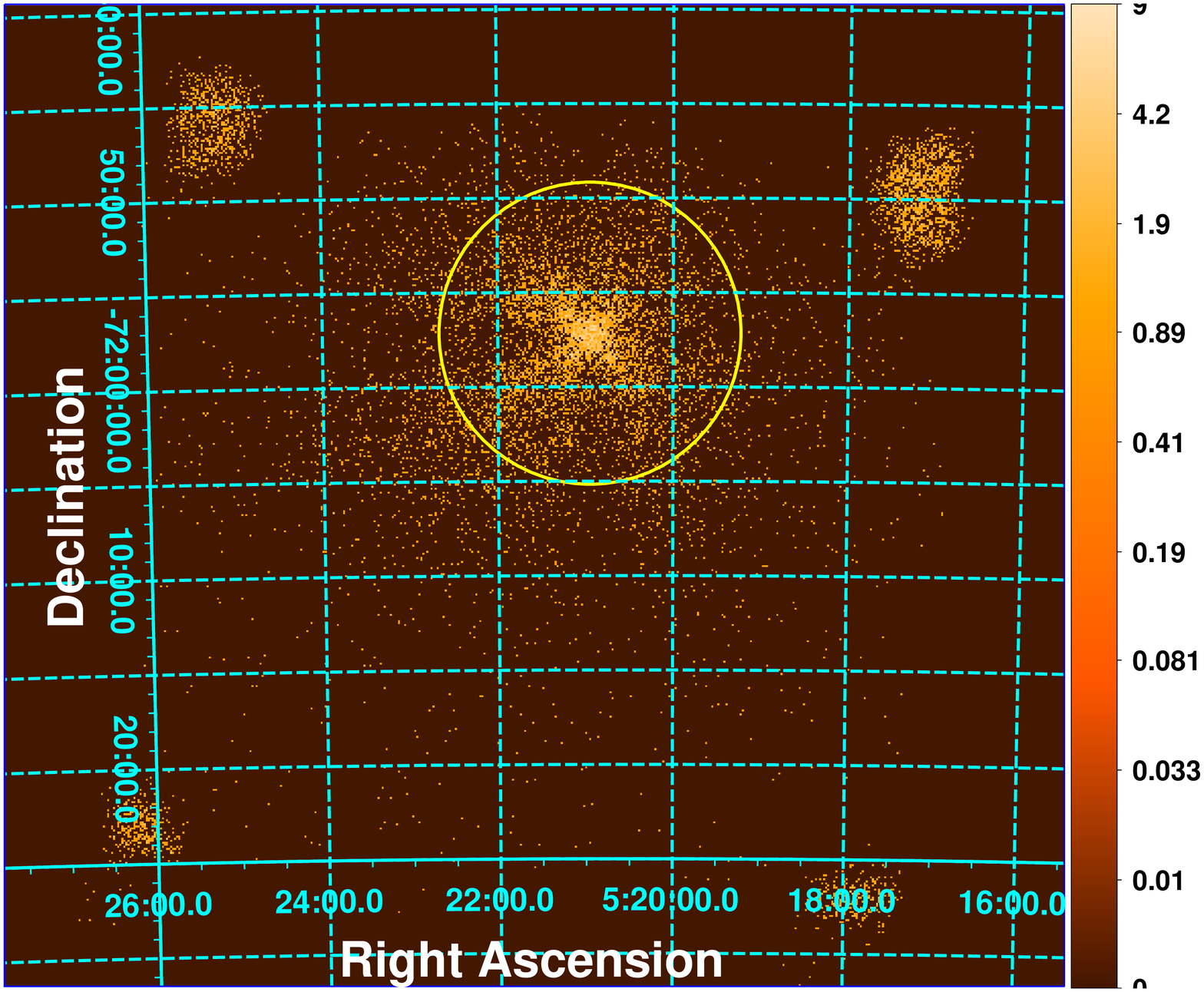}
\caption{The image of LMC X-2 as observed by {\it SXT} in June 2016. The spectra and lightcurves 
have been extracted from a circular region of 8 arcmin as shown in the figure. The four spots 
seen at the corners of the image are calibration sources ($Fe^{55}$).}
\end{figure}

We used XSELECT version 2.4d to extract the image, lightcurves
and spectrum from the {\it SXT} level2 event files provided by the
instrument team. We used a circular region of 8 arcmin centered
at the source position to extract the source  spectrum and
lightcurves. In Figure 1, we show the {\it SXT} image of the LMC X-2 for the Orbit number 3977. 
We created the instrument ARF using the tool {\it sxtmkarf} provided by instrument team 
\footnote{http://www.tifr.res.in/$\sim$astrosat\_sxt/dataanalysis.html},
which also takes care of the off-pointing correction. We utilized the
latest version of the software ``LaxpcSoft" provided by the {\it LAXPC} team
\footnote{http://www.tifr.res.in/$\sim$astrosat\_laxpc/LaxpcSoft.html}
to analyze the {\it LAXPC} data and followed the procedure described there 
(see also Agrawal et al. 2018; Sreehari et al. 2019).

\section{Data Analysis}

\subsection{Lightcurve and Z-track}

We used the top layer LAXPC10 event data to construct the lightcurve. Since above 20 keV 
the intensity is dominated by the contribution
from the background, we created the lightcurve in the energy band of $3.0-20.0$
keV. In Figure 2, we show the background subtracted binned lightcurve in
the energy range of $3.0-20.0$ keV. The binsize used here is 256 seconds. The
source exhibited a flare during our observations. The background subtracted lightcurve 
created using {\it SXT} data is shown in Figure 3.

\begin{figure}
\centering
\includegraphics[width=0.35\textwidth,angle=-90]{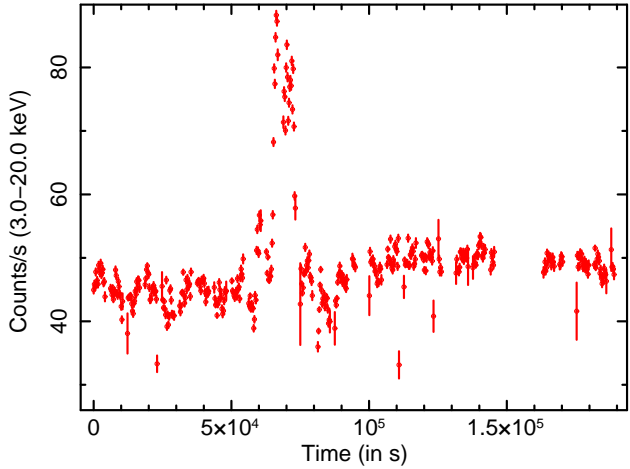}
\caption{The background subtracted lightcurve of LMC X-2 in the energy range $3.0-20.0$ keV. 
The lightcurve is created using the LAXPC10 data. The time binsize used for the lightcurve 
is 256~s.}
\end{figure}

\begin{figure}
\centering
\includegraphics[width=0.35\textwidth,angle=-90]{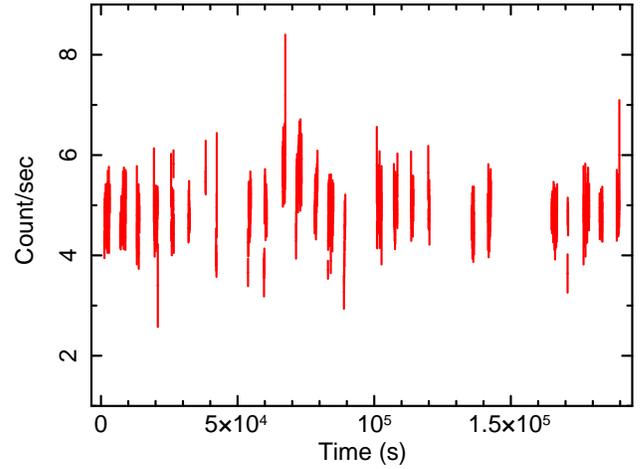}
\caption{The background subtracted {\it SXT} lightcurve of LMC X-2 in the energy range 
$0.3-8.0$ keV. The time binsize used for the lightcurve is 64~s.}
\end{figure}

Hardness Intensity Diagram (HID) was created using the background subtracted lightcurves.  
We defined the overall colours as the ratio of count rates in the energy bands $6.5-18.5$ keV and 
$3.0-6.5$ keV, whereas intensity was defined as the total count rates in the $3.0-18.5$ keV
energy band. We plotted the overall colours against the source
intensity to construct the HID. The HID is shown in Figure 4. We used a
binsize of 1024~s to create the HID. The large time binsize has been used
here to separate the three branches of the Z-pattern. All the three branches of the 
Z-track can be identified clearly in the HID (Figure 4). The HID shows remarkable similarity 
with that obtained using the {\it RXTE} data \citep{Smale03,agr09}.
 
In order to study the evolution of the broadband spectral and temporal
behaviour along the complete Z-track, we divided this track into 7
segments. We divided the horizontal branch (HB) into  two sections `HB1'
and `HB2'. The normal branch (NB) was also divided into two sections
namely `NB1', `NB2'. The points close to the lower-most NB and the bottom
part of the FB are part of the NB-FB vertex (NBV). The rest of the FB
was divided into two sections `FB1' and `FB2'.

\begin{figure}
\centering
\includegraphics[width=0.35\textwidth,angle=-90]{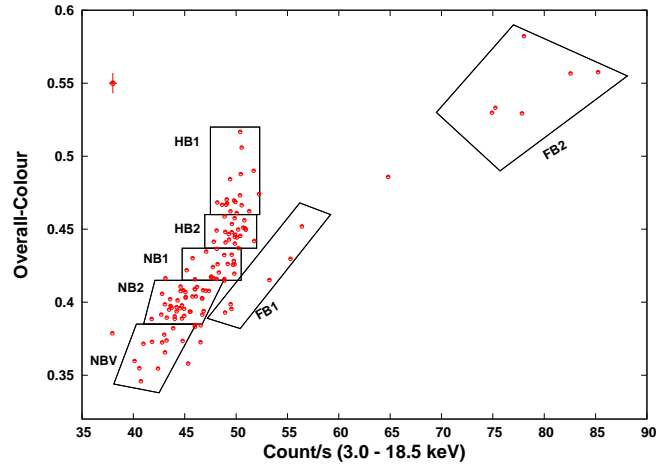}
\caption{Hardness Intensity Diagram (HID) of LMC X-2 created using the LAXPC10 data. To 
create the HID, we have used 1024~s binsize. Different sections of the HID have been marked. 
A typical error bar on the data points is shown in the top-left corner. See text for details.}
\end{figure}

\subsection{Spectral Analysis} 

We used the {\it LAXPC10} top layer data to create the source and background
spectra for different sections of the HID. The latest {\it SXT} and {\it LAXPC}
response matrix files provided by the instrument team were used  for the
spectral analysis. The sky background spectrum provided by the {\it SXT} team was
used for the background subtraction. We used combined {\it SXT} ($0.5-5.5$ keV)
and LAXPC10 ($3.0-20.0$ keV) data for spectral analysis. We restricted
our analysis to these energy ranges because there is not enough source flux
above 5.5 keV for the {\it SXT} and above 20 keV for the {\it LAXPC}. The combined
spectra in the energy range $0.5-20$ keV were fitted with the XSPEC version 12.9.1. 
We grouped the {\it SXT} data to give a minimum of 25 counts/bin. All the errors were computed 
using $\Delta \chi^2=1.0$ (68\% confidence level). We added 1\% systematics during fitting 
to account for the uncertainty in the response matrix. 

According to the Dickey and Lockman survey, the galactic $N_H$ in the direction of LMC X-2 is  0.063 $\times$ 10$^{22}$ $cm^{-2}$ \citep{Dickey90}. However, $N_H$ calculated using the more recent surveys like Leiden/Argentine/Bonn (LAB) \citep{Kalberla05} and HI4PI \citep{Bailin16} is 0.15 $\times$ 10$^{22}$ $cm^{-2}$. To calculate the $N_H$,  we used the nH calculator tool provided by HEASARC, NASA. We used the multiplicative model {\it Tbabs} of XSPEC to account for the galactic absorption.

First, we fitted the spectra with a simple {\it cutoffpl+bbodyrad}
model. This model has been used to describe the spectra of Z-sources
(Birmingham model; see \citealt{Jackson09}; \citealt{balu10}). We found
that for all parts of the Z-track photon index of the cutoff powerlaw
was $<$ 1.  Hence {\it  cutoff-powerlaw} is not consistent with the
Comptonization model. We refer to this model as {\bf Model 1}. We tried
using other complex Comptonization models such as {\it nthComp}.  The {\it
nthComp} model has provision to select the blackbody or disc blackbody as
the input seed  photon spectrum \citep{Zdz96}. Yet, another Comptonization
model such as {\it compTT} which has been used previously to model the
spectrum of this source \citep{lava08,agr09} assumes a Wien spectrum for
the seed photons \citep{Titar94}. The model provided  statistically good
fit to the spectra of all the sections of the Z-track. We refer to this
model as {\bf Model 2}. We also tried a combination of disc blackbody
({\it diskbb}) and {\it nthComp}, as well as a combination of blackbody
({\it bbody}) and {\it nthComp} model. We noticed that addition of a
{\it blackbody} or a {\it diskbb} component to the {\it nthComp} did
not improve the fit. This suggests that, probably the disc is obscured
by a Comptonized corona. The seed photons for the Comptonization may
come from the obscured inner accretion disc or from the surface of the
central source. Here, we assumed that the seed photons comes from the
accretion disc. However, it is possible that the blackbody emission
from the surface of the neutron star may also supply the seed photons
for the Comptonization.
\begin{figure}
\includegraphics[width=0.35\textwidth,angle=-90]{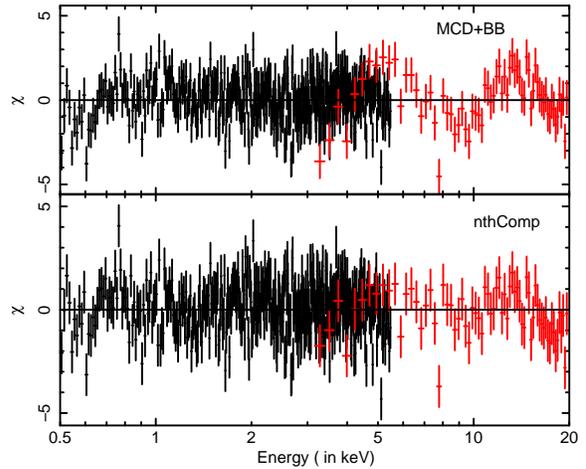}
\caption{The residual in unit of sigma for {\it diskbb+bbodyrad} (MCD+BB) model (top panel) 
and {\it nthComp} model (bottom panel). From figure, it is clear that the {\it nthComp} model 
provides a better fit compared to the {\it diskbb+bbodyrad} model.}
\end{figure}

\begin{figure}
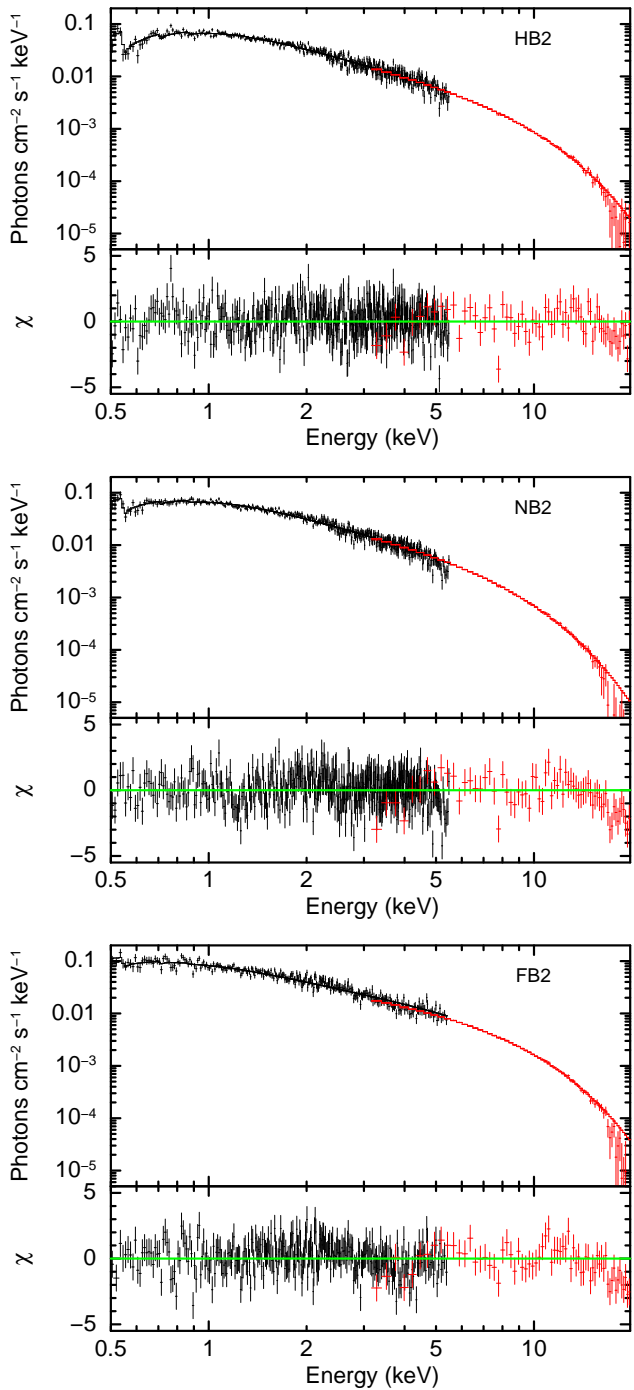

\includegraphics[width=0.35\textwidth,angle=-90]{hb2-combined-nthcomp-seedDISK.eps}
\includegraphics[width=0.35\textwidth,angle=-90]{nb2-nthcomp-combined-seedDISK.eps}
\includegraphics[width=0.35\textwidth,angle=-90]{fb2-nthcomp-combined-seedDisk.eps}
\caption{The unfolded spectra at different parts (HB2, NB2, FB2) of the Z-track along with 
the residual (bottom panels) in the units of sigma.}
\end{figure}

We also fitted the data with {\it diskbb+bbodyrad} model, which has been previously used to model the spectra of this source \citep{lava08} and the other
LMXBs \citep{Mit84}. The above combination provided a poor fit compared
to the {\it nthComp} model for the HB and NB sections (see Table 2 and
Table 3). However, {\it diskbb+bbodyrad} and {\it nthComp} model gives
similar reduced $\chi^2$ ($\chi^2/dof$) in the FB. We refer to this
model as {\bf Model 3}.  In Figure 5, we show the residual resulted by
fitting the spectra of the section HB2 with {\bf Model 2} and {\bf Model
3}. From the residual, it is clear that {\bf Model 2} (bottom panel of
Figure 5) is better compared to {\bf Model 3} (top panel of Figure 5).
While fitting the spectra with these models, we did not find any residual
around $\sim$ $5-7$ keV range, suggesting that the iron line (Fe-K)
feature is absent in this source during our observations.

\begin{figure}
\includegraphics[width=0.45\textwidth,angle=-90]{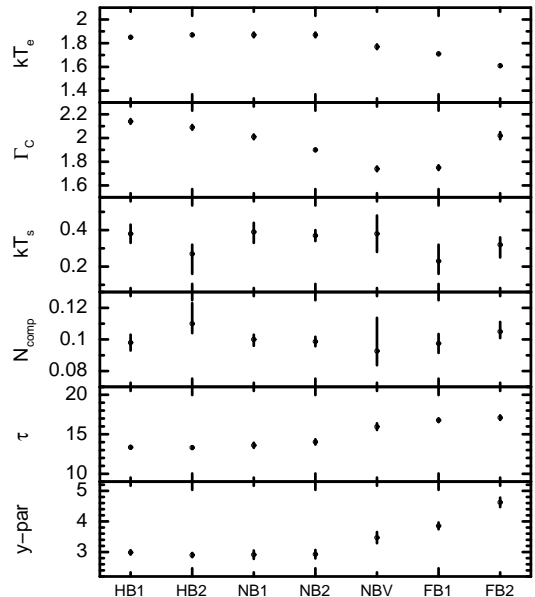}
\caption{The evolution of the best fit spectral parameters of {\bf Model 2} along the Z-track. 
$y-par$ (Comptonization parameter) and $\tau$ (optical depth) are derived parameters 
(see text for details).}
\end{figure}

\begin{figure}
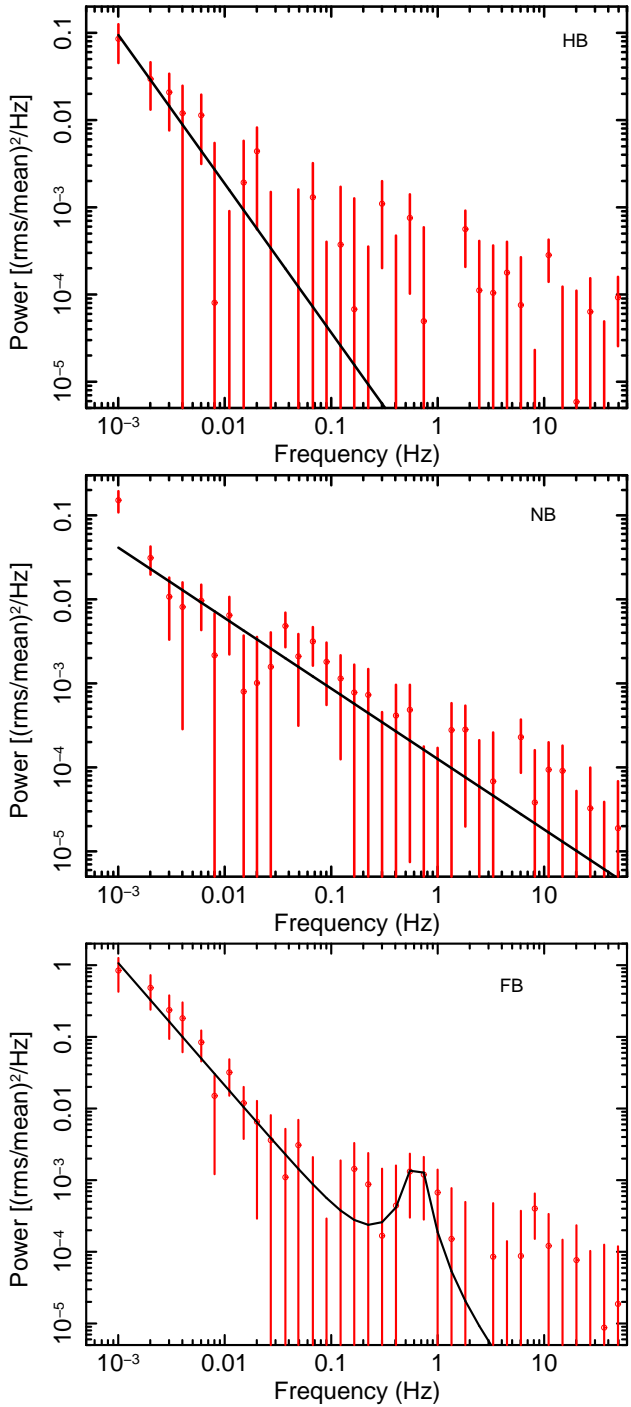

\includegraphics[width=0.35\textwidth,angle=-90]{hb-pds-4ms-combined.eps}
\includegraphics[width=0.35\textwidth,angle=-90]{nb-pds-combined-4ms.eps}
\includegraphics[width=0.35\textwidth,angle=-90]{fb-pds-combined-4ms.eps}
\caption{Power density spectra (PDS) for different branches (HB, NB, FB) of the Z-track 
along with the best fit Model. Marginal detection of QPO-like feature is seen in FB. See, text
for details.}
\end{figure}

\subsection{Timing Analysis}

We used the LAXPC lightcurves in the energy range $3.0-20.0$ keV to create the PDS. The lightcurves with binsize 4 ms were divided into intervals of  262144 bins, which allowed to study the nature of timing variabilities in the range of $0.001-125.0$ Hz. PDS were created for each interval and those belonging to the same section of the HID are averaged. We rebinned the PDS in the frequency space by a factor of 1.3. The binned PDS were normalized to  the fractional rms spectra (in units of $(rms/mean)^2 Hz^{-1}$) and an appropriate Poisson noise was subtracted \citep{Zhang95,agr18}. To get a better statistics, we merged the PDS belonging to the same branch. While averaging the PDS for the NB branch, we assigned NBV data points as a part of the NB.  PDS  of all these three branches  can be fitted with a simple power-law ($A\nu^{-\alpha}$). While fitting the PDS of FB with a power-law model, we found a signature of a weak QPO-like feature at $\sim$  $0.6-0.7$ Hz. Fitting this feature with an extra Lorentzian component improved the fit. The reduced $\chi^2$ decreased from 16.2/32 to 11.5/30.
\begin{table*}
\caption{The best fit spectral parameters for {\bf Model 1} ({\it Tbabs*(cutoffpl+bbodyrad)}).  
$\Gamma$ is photon index and $E_C$ is cutoff energy. $K$ is the normalization of the cutoff 
powerlaw at 1 keV and in units of $photons/s/cm^2/keV$. $N_{BB}$ is the normalization of the 
blackbody component and $kT_{BB}$ is its temperature. $N_H$ is in units of $10^{22}$ cm$^{-2}$.}
\begin{tabular}{cccccccc}
\hline
Parameters &   HB1 & HB2 & NB1 & NB2 & NBV & FB1 & FB2 \\
\hline
\hline
$N_H$ &        0.065(fix) & 0.10$\pm$0.01 & 0.065(fix) & 0.065(fix)&0.065(fix)&0.065(fix)&0.065(fix)\\
$\Gamma$ & 0.27$\pm$0.06 & 0.41$\pm$0.06 &0.30$\pm$0.06&0.21$\pm$0.05&0.15$\pm$0.07&-0.16$\pm$0.09&-0.26$\pm$0.05\\
$E_C$ (keV)  &2.94$\pm$0.08 & 2.97$\pm$0.08&2.75$\pm$0.06&2.54$\pm$0.05& 2.38$\pm$0.06&2.20$\pm$0.06&2.58$\pm$0.05 \\
$K$ ($\times10^{-2})$ &4.85$\pm$0.4 &6.28$\pm$0.5 & 5.86$\pm$0.3 &5.45$\pm$0.2&6.76$\pm$0.5 &465.6$\pm$0.4 & 458$\pm$0.3 \\
$kT_{BB}$ (keV) & 0.42$\pm$0.01&0.37$\pm$0.01 &0.39$\pm$0.01& 0.39$\pm$0.06&0.31$\pm$0.02&0.33$\pm$0.01 &0.36$\pm$0.01\\
$N_{BB}$ &  350.0$\pm$32& 515.0$\pm$50.5 & 390.1$\pm$37.5&445.2$\pm$32.1&1172.5$\pm$240.7&856.4$\pm$92.5&893.8$\pm$65.6\\
$\chi^2/dof$ & 422/327 & 667/488& 605/378&675/491&216/168&324/299&526/368\\
\hline
\end{tabular}
\end{table*}

\begin{table*}
\caption{The best fit spectral parameters for {\bf Model 2} ({\it Tbabs*nthComp}). $kT_{e}$ and  
$kT_s$ are the electron temperature and seed photon temperature respectively. 
$\Gamma_C$ and $N_{Comp}$ are photon index and the normalization of {\it nthComp} component
respectively. $\tau$ is the optical depth and $y$-par is the
Comptonization parameter. $F_{Comp}$ is the $0.1-50$ keV unabsorbed
Comptonization flux in units of ergs/s/cm$^2$. $N_H$ is the
neutral hydrogen column density in untis of 10$^{22}$ cm$^{-2}$. $L_X$ is $0.1 - 50$ keV 
unabsorbed luminosity in units of ergs/s.}
\begin{tabular}{cccccccc}
\hline
\hline
Parameters & HB1        & HB2 &        NB1      & NB2      & NBV& FB1 & FB2\\
\hline
$N_H$       &0.17$\pm$0.02 & 0.22$\pm$0.02 & 0.16$\pm$0.02 & 0.16$\pm$0.01&0.08$^{+0.06}_{-0.03}$ & 0.13$\pm$0.3& 0.09$\pm$0.01\\
$kT_e$ (keV)&    2.14$\pm$0.02 &2.09$\pm$0.02& 2.01$\pm$0.02 &1.90$\pm$0.01 & 1.74$\pm$0.02&1.75$\pm$0.02 & 2.02$\pm$0.02 \\
$\Gamma_C$& 1.85$\pm$0.01 & 1.87$\pm$0.01 & 1.87$\pm$0.01 & 1.87$\pm$0.02 & 1.77$\pm$0.02& 1.71$\pm$0.01 & 1.61$\pm$0.01\\
$kT_s$ (keV) &0.38$\pm$0.05 &0.27$^{+0.05}_{-0.11}$&0.39$\pm$0.05 & 0.37$\pm$0.03 & 0.38$\pm$0.1 & 0.23$^{+0.09}_{-0.07}$ &0.32$\pm$0.06\\
$N_{Comp}( \times 10^{-2})$& 9.81$\pm$0.5 & 11.05$^{+1.3}_{-0.6}$ & 10.5$\pm$0.3 & 9.87$\pm$0.3 & 9.27$^{+2.1}_{-0.9}$ & 9.75$\pm$0.6 & 10.5$\pm$0.6\\
$F_{comp}$ $(\times 10^{-10}$) &3.89$\pm$0.09 & 3.98$\pm$0.09 & 3.80$\pm$0.17 & 3.63$\pm$0.08 & 3.46$\pm$0.08 & 3.89$\pm$0.09 & 6.02$\pm$0.14\\
\hline
$R_W$ (km) & 102$\pm$26.5& 206.5$\pm$110.4& 97.2$\pm$24.3& 149.5$\pm$25.5& 89.5$\pm$36.5&253.5$\pm$110.5&150.2$\pm$55.4\\
$\tau$   &13.35$\pm$0.16 & 13.31$\pm$0.15 & 13.65$\pm$0.3 &  14.05$\pm$0.31&15.97$\pm$0.41 & 16.77$\pm$0.24 & 17.15$\pm$0.28 \\
$y-par$&2.98$\pm$0.06 & 2.91$\pm$0.06 & 2.91$\pm$0.12 & 2.92$\pm$0.13 & 3.47$\pm$0.15 & 3.85$\pm$0.11 &4.62$\pm$0.14\\
$L_X$ $(\times 10^{38})$ & 1.15$\pm$0.02 & 1.18$\pm$0.03 & 1.13$\pm$0.05 & 1.08$\pm$ 0.02 & 1.03$\pm$0.02 & 1.15$\pm$0.03 & 1.79$\pm$0.04\\
\hline
$\chi^2/dof$ &   415/327 & 663/489 &604/378 &675/491 & 197/168 &309/299 & 484/368\\
\hline
\hline
\end{tabular}
\end{table*}

\begin{table*}
\caption{The best fit spectral parameters for {\bf Model 3} ({\it Tbabs*(diskbb + bbodyrad}).  
$kT_{bb}$ is the blackbody temperature and $kT_{in}$ is the inner disc temperature. $N_{bb}$ 
and $N_{dbb}$ are the normalization of the {\it bbodyrad} and {\it diskbb} components 
respectively. $N_H$ is in units of 10$^{22}$ cm$^{-2}$.}
\begin{tabular}{cccccccc}
\hline
\hline
Parameters & HB1        & HB2 &        NB1      & NB2      & NBV& FB1 & FB2\\
\hline
$N_H$   &{\bf 0.09$\pm$0.01} & 0.09$\pm$0.01 & 0.10$\pm$0.006 & 0.063(fixed) & 0.07$\pm$0.01& 0.063(fixed) & 0.063(fix)\\ 
$kT_{in}$ (keV)   &0.97$\pm$0.02 & 0.88$\pm$0.01 & 0.93$\pm$0.02 & 0.84$\pm$0.02 & 0.87$\pm$0.02 & 0.81$\pm$0.02 & 0.92$\pm$0.01\\
$N_{dbb}$    &19.42$\pm$1.92 & 29.71$\pm$1.85 & 22.91$\pm$2.07 & 32.29$\pm$2.04 & 28.48$\pm$3.36 & 35.27$\pm$4.12&27.35$\pm$1.65\\
$kT_{bb}$ (keV)  & 1.80$\pm$0.02 & 1.72$\pm$0.01 & 1.71$\pm$0.01 & 1.60$\pm$0.01 & 1.54$\pm$0.01 & 1.51$\pm$0.01 & 1.79$\pm$0.01\\
$N_{bb}$ & 2.21$\pm$0.11 & 2.70$\pm$0.10 & 2.71$\pm$0.13 & 3.32$\pm$0.13 & 4.35$\pm$0.24 & 4.87$\pm$0.26 & 4.56$\pm$0.15\\
\hline
$R_{dbb}\sqrt{\cos\theta}$ (km) &22.03$\pm$1.08 & 27.25$\pm$0.85 & 23.95$\pm$1.08& 28.42$\pm$0.89 & 26.68$\pm$1.58& 29.70$\pm$1.72& 26.15$\pm$0.78\\
$R_{bb}$ (km) & 7.43$\pm$0.18 & 8.21$\pm$0.15 & 8.23$\pm$0.19 & 9.09$\pm$0.18 & 10.42$\pm$0.28 & 11.03$\pm$0.29 & 10.67$\pm$0.17\\
\hline
$\chi^2$/dof&440/327  & 759/489 & 651/378 & 705/491 & 211/168 & 307/300 & 481/369\\
\hline
\hline
\end{tabular}
\end{table*}

\begin{table*}
\caption{The best fit parameters obtained by fitting the PDS of different branches of the 
Z-track. $\alpha$ is the index and $A$ is the normalization of the power-law component. 
Pow-rms is the total rms of the power-law component in the frequency range $0.001-50$ Hz and 
QPO-rms is the rms of the 
QPO feature in the frequency range $0.001-50$ Hz.}
\begin{tabular}{cccc}
\hline
Parameters& HB & NB & FB \\
\hline
\hline
$\alpha$ & 1.70$\pm$0.22 &0.83$\pm$0.12 & 1.71$\pm$0.11\\
$A$($\times 10^{-5}$) & 0.078$\pm$0.03 &13.4$\pm$2.5 &0.87$\pm$0.3\\
Pow-rms (\%)  &1.21$\pm$0.82 & 3.61$\pm$1.62 & 3.97$\pm$1.58 \\
$\nu_c$ (Hz)    &&&0.65$\pm$0.07  \\
$FWHM$ (Hz)   &&&0.2 (fixed)  \\
LN ($\times$ 10$^{-3}$) &&& 2.48$\pm$0.84 \\
QPO-rms (\%) & & & 2.71$\pm$0.52\\
F-test       & & & 5.8 $\times$ 10$^{-3}$\\
\hline
$\chi^2$/dof &  27.1/32 &26.2/32 &11.5/30\\ 
\hline
\end{tabular}
\end{table*}

\section{Results}

\subsection{Spectral Behaviour} 

The broadband spectra ($0.5-20.0$ keV) along the Z-track can be fitted
with an absorbed Comptonization model ({\bf Model 2}). We also tried
the combination of emission from the multicolour disc (described by {\it
diskbb} in XSPEC) and blackbody component ({\bf Model 3}). We found
that {\bf Model 3} is not satisfactory for describing the spectra of
the HB and the NB (see Table 3). The spectral parameters and reduced
$\chi^2$ of {\bf Model 2} and {\bf Model 3} are given in Table 2 and Table
3 respectively. {\bf Model 3} gave the inner disc temperature in the
range of $\sim$ $0.8-1.0$ keV and black body temperature in the range
of $1.5-1.8$ keV. In Figure 6, we show the spectra of the HB2, NB2 and FB2
fitted with {\bf Model 2}.
The $N_H$ values derived using {\bf Model 2} lie  in the range of  $(0.1-0.2)$ $\times$ $10^{22} cm^{-2}$. The derived $N_H$ values  in the direction of LMC X-2 are close to the results of recent surveys \citep{Kalberla05,Bailin16}. Moreover, systematics in the data and model can also affect the best-fit values of $N_H$.

Fitting the spectra with {\bf Model 2} gave the electron temperature
in the range of $1.7-2.1$ keV (see Table 2). The electron temperature decreased as the
source moved from the HB to the NB and then again increased as it moved up
along the FB. The photon index of the Comptonized emission was around $\sim$
1.85 from the top-left HB to the lower NB. Then as the source moved down
the NB-FB vertex (NBV) and then up in the FB the photon index decreased
or the source spectrum became harder. We also computed the optical
depth by formula given in \cite{Zdz96} and the Comptonization parameter
using a relation $y=\frac{4 kT_e}{m_e c^2} max(\tau, \tau^2)$. The
optical depth did not show significant change from HB1 to NB2 and
remained in the range of ~ $13.2-14.3$. As the source moved to the vertex (NBV), the optical depth increased from 14.05$\pm$0.31 to 15.97$\pm$0.41 and
continued to increase along the FB. The Comptonization parameter $y$
decreased from $\sim$ 2.98 to to $\sim$ 2.92 as the source moved down the
HB and then remained constant in the NB (NB1 and NB2). The $y$ parameter
increased from 3.47$\pm$0.15 to 4.62$\pm$0.14 with the movement of the
source from the vertex (NBV) to the FB2. Since the distance ($\sim$
50 kpc) to the source is known with a better accuracy \citep{freed01},
the uncertainty in the intrinsic luminosity of the source is also much
less compared to other Z-sources. The luminosity of the source showed a
systematic decrease as the source moved from the upper NB to the vertex
(NBV), and as the source further moved up the FB the luminosity increased. The
luminosity varied in the range $(1.03-1.79)$ $\times 10^{38}$ ergs/s.
We show the evolution of the best fit spectral parameters of {\bf Model 2} in Figure 7.

We also computed the the radius of the seed photon emitting region using the formula 
(see \citealt{Zand99}),

\begin{equation}
R_W = 3 \times 10^4 D \frac{\sqrt{\frac{f_{bol}}{1+y}}}{kT_s^2}.
\end{equation}
The Wein radius $R_W$ was found to be in the range of $70-310$ km and
did not show any clear correlation with the position on the Z-track
(see Table 2). Here, $D$ is the distance to the source in kpc,
$y$ is the Comptonization parameter and $f_{bol}$ is the bolometric
($0.1-50$ keV) flux of the Comptonized component. 

 We also derive the inner disc radius and the blackbody radius using the parameters of {\bf Model 3}. The inner disc radius $R_{dbb}\sqrt{\cos\theta}$ is given by $\sqrt{N_{dbb}} D_{10}$, where $\theta$ is the inclination angle and $D_{10}$ is the distance to the source in the unit of 10 kpc. Similarly, the radius of blackbody emitting region is given by $\sqrt{N_{bb}} D_{10}$. The inner disc radius is found to be in the range of $20-30$ km and the blackbody radius varies in the range of $7-11$ km.  

\subsection{Power-spectral Properties}
We detected very low-frequency noise (VLFN) in all the three branches
(HB, NB and FB). For the NB, we found  index $\alpha$ = 0.83$\pm$0.11 and
integrated rms ($0.001-50$ Hz) = 3.61$\pm$1.62\%. The VLFN in the HB has
integrated rms = 1.21$\pm$0.82\% and that in the FB has integrated rms
= 3.97$\pm$1.58\%. We found a marginal evidence of QPO at 0.65$\pm$0.07
Hz in the FB with significance $\sim$ 2.8~$\sigma$ and integrated rms ($0.001- 50$ Hz) = 2.71$\pm$0.52\%. The best fit PDS
model parameters for HB, NB and FB are presented in Table 4. Figure 8
shows the PDS for these three branches along with the best fit model.
\section{Discussion}

The bright low-mass X-ray binary LMC X-2 traced a complete Z-track during
the $\sim$ 140 ks {\it LAXPC} observations. We also note that the range of the
overall color values ($0.3-0.6)$  during our observations is similar to
that observed during the {\it RXTE} observations \citep{Smale03,agr09}.
The track in the HID is almost identical during the {\it RXTE} and the present
{\it AstroSat} observations. We also note that the source spent $\sim$
12 ks or 8\% time in the flaring state. Using these observations,
we studied the evolution of the broadband ($0.5-20$ keV) spectral
parameters along the complete Z-track by combining the {\it SXT} and {\it LAXPC}
data, which is a first such study. Previously, spectral evolution of
the source has been investigated along the Z-track using the data from
the {\it RXTE-PCA} in the energy range of $3.0-20.0$ keV \citep{agr09}.

Here, we have shown that the $0.5-20.0$ keV spectra of the source can be
described by a simple absorbed Comptonization model ({\bf Model 2}). We
find that the Comptonizing region is cool and optically thick with the
electron temperature in the range of $\sim$ $1.7-2.1$ keV and the optical
depth in the range of $\sim$ $13.2-17.5$. The {\it RXTE-PCA} spectra
in the $3.0-20.0$ keV energy band, fitted with {\it compTT} model provided
a similar optical depth ($\tau \sim 13.0-18.0$) but a slightly higher
electron temperature ($kT_e \sim 1.95-2.7$ keV) \citep{agr09}.

Most importantly, we do not require an extra disc emission component or a
blackbody component to fit the spectra in the $0.5-20.0$ keV range unlike
{\it Suzaku} observations \citep{agr09}. It is worth mentioning that the
$0.3-10.0$ keV {\it XMM-Newton} spectrum of the source was also modeled
using a pure Comptonized emission \citep{lava08}. The possible explanation
for the absence of the soft component is that the temperature of the soft
component may be well below the lower energy bound of the instrument,
considered while fitting.  The seed photon temperature $kT_{s}$ in our
case is $\sim$ $0.2-0.4$ keV. However, a significantly lower value of
$kT_s$ ($\sim$ 40 eV) has been reported by \cite{lava08} using the {\it
XMM-Newton} observation. They also calculated the radius of seed photons
emitting region and it was found to be very high ($R_W \sim 10^4$ km) and
which lead to conclusion that Comptonization model is not physically
acceptable. In our case, the radius of seed photon emitting region is
$70-310$ km and hence the Comptonization model ({\it nthComp}) is an
appropriate to represent the spectra of the source along the Z-track.
  
We have studied the evolution of the spectral parameters along the
Z-tarck. Though variations in these parameters are subtle, they provide
an important probe to understand the movement of the source along the
Z-pattern. Our findings suggest that the optical depth of the corona
increases as the source moves from the NB-FB vertex (NBV) to the upper
FB. We also note that the Compton $y$ parameter also increases along the
FB. This trend is also seen in other Z-sources (GX 349+2: \citealt{agr03},
GX 17+2: \citealt{agr20}) and in the previous observations of LMC X-2
\citep{agr09}. The observed variations in the spectral parameters ($\tau$
and $y$) suggest that most probably at the vertex of the NB and the FB,
the accretion rate crosses the Eddington limit and a strong radiation
pressure drives a fraction of the disc material into a hot central
corona. The above scenario has been discussed by \cite{agr03,agr20} to
explain the increase of the optical depth along the FB in the  Z-sources
GX 349+2 and GX 17+2. Hence, the motion from the vertex to the upper FB
can be understood in terms of increasing accretion rate scenario.
  
It has been observed that in some of the Z-sources the optical depth of
the Comptonized component generally decreases as the source moves from
the HB to the lower NB (GX 17+2: \citealt{disalvo00,agr20}, GX 349+2:
\citealt{agr03}, GX340+0: \citealt{Iaria06}, Cyg X-2: \citealt{disalvo02})
and the electron temperature either remains constant \citep{agr20,
disalvo00} or increases \citep{disalvo02, agr03, Iaria06}.  In the present
case, the electron temperature of the Comptonizing region decreases as
the source moves from HB to the vertex of the NB and FB. However, the
optical depth remains nearly constant upto the middle of NB (NB2). The
source exhibited a similar behaviour during the {\it RXTE} observations
\citep{agr09}. The decrease in the $kT_e$ can be explained in terms of
cooling of the corona due to increase in the seed photon supply. However,
it is also possible that a part of the coronal material cools down and
settles in the disc causing the optical depth to decrease and the
temperature of the remaining material in the corona will either not
change or will increase. The above scenario can explain the spectral
behaviour of other Z-sources.

We have derived PDS for the three branches (HB, NB and FB). In
all three branches, PDS  has the VLFN component.  \cite{Smale03}
found VLFN component with $\alpha$ $\sim$ 0.60 in the HB, $\alpha$
$\sim$ 0.90 in the NB and $\alpha \sim$ 1.33 in the FB. In our case,
the VLFN in the HB and FB is steeper compared to the previous results
\citep{Smale03}. However, the value of VLFN index in the NB is close
to that obtained by \cite{Smale03}. We also detect a weak ($\sim$
$2.8~\sigma)$ QPO at $\sim$ 0.65 Hz in the FB. The rms amplitude of the
QPO is $\sim$ 2.7\%.

\section*{Acknowledgements}
Authors thank the anonymous reviewer for providing useful suggestions
which improved the quality of the manuscript. This research has made use
of the data obtained through GT phase of {\it AstroSat} observation.
Authors thank DD, PDMSA and Director, URSC for encouragement and
continuous support to carry out this research. Authors also thank
Ravishankar B. T. of SAG for careful reading the manuscript and providing
useful comments. This work has used the data from the LAXPC Instruments
developed at TIFR, Mumbai and the LAXPC POC at TIFR is thanked for
verifying and releasing the data via the ISSDC data archive. We thank
the AstroSat Science Support Cell hosted by IUCAA and TIFR for providing
the LaxpcSoft software which we used for LAXPC data analysis.  This work
has used the data from the Soft X-ray Telescope ({\it SXT}) developed at
TIFR, Mumbai, and the {\it SXT} POC at TIFR is thanked for verifying \&
releasing the data and providing the necessary
software tools.\\

\section*{data availability}
Data underlying this article are available at {\it AstroSat}-ISSDC website
(http://astrobrowse.issdc.gov.in/astro\_archive/archive).


\begin{thebibliography}{99} 
\bibitem[\protect\citeauthoryear{Agrawal et al.}{2020}]{agr20}
Agrawal V. K., Nandi Anuj, Ramadevi M. C., 2020, Ap\&SS, 365, 41
\bibitem[\protect\citeauthoryear{Agrawal et al.}{2018}]{agr18}
Agrawal V.K., Nandi Anuj, Girish V., Ramadevi M.C., 2018, MNRAS, 477, 5437
\bibitem[\protect\citeauthoryear{Agrawal and Misra}{2009}]{agr09}
Agrawal V.K., Misra R., 2009, MNRAS, 398, 1352
\bibitem[\protect\citeauthoryear{Agrawal and Sreekumar}{2003}]{agr03}
Agrawal V.K., Sreekumar P., 2003, MNRAS, 346, 933
\bibitem[\protect\citeauthoryear{Bailin et al.}{2016}]{Bailin16}
Bailin J., Calabretta M.R., Dedes L., Ford H.A., Gibson B.K., Haud U., Janowiecki S., Kalberla P.M.W. et al., 2016, A\&A, 594, A116
\bibitem[\protect\citeauthoryear{Balucinska-Church et al.}{2010}]{balu10}
Balucinska-Church M., Gibbec A., Jackson N.K., Church M.J., 2010, A\&A,512, A9 
\bibitem[\protect\citeauthoryear{Bhulla et al.}{2019}]{Bhulla19}
Bhulla Y., Misra R., Yadav J.S. et al.,2019, RAA, 19, 114
\bibitem[\protect\citeauthoryear{Bonnet-Bidaud et al.}{1989}]{Bonnet89}
Bonnet-Bidaud J.M., Motch C, Beuermann K., Pakull M., Parmar A.N., van der Klis M., 1989, A\&A, 213, 97
\bibitem[\protect\citeauthoryear{Cackett et al.}{2010}]{Cackett10}
Cackett E.M., Miller J. M., Ballantyne D. R., Barret D., Bhattacharyya S., Boutelier M., Miller M.C., Strohmayer T. E., 2010, ApJ, 720, 205
\bibitem[\protect\citeauthoryear{Church et al.}{2012}]{Church12}
Church M.J., Gibiec A., Balucinska-Church M., Jackson N.K., 2012, A\&A, 546, A35
\bibitem[\protect\citeauthoryear{Dickey and Lockman}{1990}]{Dickey90}
Dickey J.M., Lockman F.J., ARA\&A, 1990, 28, 215
\bibitem[\protect\citeauthoryear{Di Salvo et al. }{2002}]{disalvo02}
Di Salvo T. et al., 2002, A\&A, 386, 535
\bibitem[\protect\citeauthoryear{ Di Salvo et al.}{2001}]{disalvo01}
Di Salvo T., et al., 2001,ApJ, 554, 49
\bibitem[\protect\citeauthoryear{Di Salvo et al. }{2000}]{disalvo00}
Di Salvo T., Stella L., Robba N.R., van der Klis M., Burderi L., Israel G.L, Homan J., Compana S. et al., 2000, ApJ, 544, L119
\bibitem[\protect\citeauthoryear{Done, Zycki and Smith}{2002}]{Done02}
Done C., Zycki P.T. and Smith D.A, 2002, MNRAS, 331, 453
\bibitem[\protect\citeauthoryear{Farinelli et al.}{2009}]{Fari09}
Farinelli R., Paizis A., Landi R., and Titarchuk L.,2009, A\&A, 498, 509 
\bibitem[\protect\citeauthoryear{Freedman et al. }{2001}]{freed01}
Freedman W.L., Madore B.F., Gibson B.K. et al., 2001, ApJ, 553, 47
\bibitem[\protect\citeauthoryear{Hasinger and van der Klis}{1989}]{Hasvan89}
Hasinger G., van der Klis M., 1989, A\&A, 225, 79
\bibitem[\protect\citeauthoryear{in 't Zand et al.}{1999}]{Zand99}
in 't Zand, J.J.M., Verbunt F., Strohmayer T.E., Bazzano A., Cocchi M., Heise J., van Kerkwijk M.H., Muller J.M., et al., 1999, A\&A, 345, 100
\bibitem[\protect\citeauthoryear{Iaria et al.}{2006}]{Iaria06}
Iaria R., Lavagetto G., Di Salvo T., D' Ai A., Burderi L., Stella L., Robba N. R., 2006, Chin. J. Astron. Astrophys., 6, 257
\bibitem[\protect\citeauthoryear{Jackson et al.}{2009}]{Jackson09}
Jackson N.K., Church M.J., Balucinska-Church M., 2009, A\&A, 494,1059
\bibitem[\protect\citeauthoryear{Johnston et al.}{1979}]{Johnston79}
Johnston M.D., Bradt H.V., Doxsey R.E., 1979, ApJ, 233 514
\bibitem[\protect\citeauthoryear{Kalberla et al.}{2005}]{Kalberla05}
Kalberla P.M.W., Burton W.B., Hartmann D., Arnal E.M., Bajaja E., Morras R., Poppel W.G.L, 2005, A\&A 440, 775
\bibitem[\protect\citeauthoryear{Lavagetto et al.}{2008}]{lava08}
Lavagetto G., Iaria R., D'Ai A., Di Salvo T., Robba N. R., 2008,A\&A, 478,181
\bibitem[\protect\citeauthoryear{Lin et al.}{2012}]{Lin12}
Lin D., Remillard R.A., Homan J., Barret D., 2012, ApJ, 756, 34
\bibitem[\protect\citeauthoryear{Markert and Clark}{1975}]{Markert75}
Markert T.H., Clark G.W., 1975, ApJ, 196, L55
\bibitem[\protect\citeauthoryear{Pakull}{1978}]{pakul78}
Pakull M., 1978, IAU Circular, 3313
\bibitem[\protect\citeauthoryear{Singh et al.}{2016}]{singh16}
Singh, K.P, Stewart G.C., Chandra S., Mukerjee K., Kotak S., Beardmore, A.P., Chitnis V., Dewangan G.C., et al., 2016, SPIE, 99051E, 10
\bibitem[\protect\citeauthoryear{Smale and Kuulkers}{2000}]{Smale00}
Smale A.P., Kuulkers E., 2000, ApJ 528, 702
\bibitem[\protect\citeauthoryear{Smale et al. }{2003}]{Smale03}
Smale A.P., Homan J., Kuulkers E., 2003, ApJ, 590, 1035
\bibitem[\protect\citeauthoryear{Mitsuda et al.}{1984}]{Mit84}
Mitsuda K., et al., 1984, PASJ, 36, 741
\bibitem[\protect\citeauthoryear{Titarchuk}{1994}]{Titar94}
Titarchuk L., 1994, ApJ, 434,570
\bibitem[\protect\citeauthoryear{Vadawale et al.}{2016}]{vadawale16}
Vadawale, S.V., Rao A.R., Bhattacharya D., Bhalerao Varun B., Dewangan G.C., Vibhute A.M., Mithun N.P.S., Chattopadhyay T., et al, 2016, SPIE, 9905, 11   
\bibitem[\protect\citeauthoryear{van der Klis }{2000}]{vander00}
van der Klis M., 2000, ARA\&A, 38, 717
\bibitem[\protect\citeauthoryear{Yadav et al.}{2016}]{Yadav16}
Yadav J.S., Agrawal P.C., Antia H.M., Chauhan Jai Verdhan, Dedhia Dhiraj, Katoch Tilak, Madhwani P., Manchanda R.K, et al., 2016, SPIE, 9905, 15
\bibitem[\protect\citeauthoryear{Zdziarski et al.}{1996}]{Zdz96}
Zdziarski A. A., Johnson W.N., Magdziarz P., 1996, MNRAS, 283, 193
\bibitem[\protect\citeauthoryear{Zhang et al.}{1995}]{Zhang95}
Zhang W., Jahoda K., Swank J.H., Morgan E.H. and Giles A.B., 1995, ApJ, 449, 930
\end{thebibliography}
\end{document}